\documentclass[conference]{IEEEtran}
\IEEEoverridecommandlockouts
\usepackage{amsmath}
\usepackage{graphicx}
\usepackage{amssymb}
\usepackage{cases}
\usepackage{cite}
\usepackage[ruled,vlined,lined,ruled,linesnumbered]{algorithm2e}
\usepackage{relsize}
\usepackage[nodisplayskipstretch]{setspace}

\begin{document}
\title{Recent Advances in Joint Wireless Energy and Information Transfer}
\IEEEspecialpapernotice{(Invited Paper)}
\author{Suzhi~Bi$^*$, Chin Keong Ho$^\dag$, and Rui Zhang$^*$$^\dag$\\
        $^*$ECE Department, National University of Singapore, Singapore. E-mail:\{bsz,~elezhang\}@nus.edu.sg\\ $^\dag$Institute for Infocomm Research, A-STAR, Singapore. E-mail: hock@i2r.a-star.edu.sg \vspace{-2ex}}
\maketitle

\vspace{-1.8cm}
\begin{abstract}
In this paper, we provide an overview of the recent advances in microwave-enabled wireless energy transfer (WET) technologies and their applications in wireless communications. Specifically, we divide our discussions into three parts. First, we introduce the state-of-the-art WET technologies and the signal processing techniques to maximize the \emph{energy transfer efficiency}. Then, we discuss an interesting paradigm named \emph{simultaneous wireless information and power transfer} (SWIPT), where energy and information are jointly transmitted using the same radio waveform. At last, we review the recent progress in \emph{wireless powered communication networks} (WPCN), where wireless devices communicate using the power harvested by means of WET. Extensions and future directions are also discussed in each of these areas.
\end{abstract}

\section{Introduction}
The proliferation of wireless communication technologies, e.g., cellular networks and wireless sensor networks (WSNs), has imposed higher requirement for the quality of power supply to wireless devices. Conventionally, wireless devices are powered by batteries, which have to be replaced/recharged manually once the energy is depleted. In practice, frequent battery replacement could be inconvenient (e.g., replace sensor batteries in a large-scale WSN) or even infeasible for some applications (e.g., implanted medical devices). Alternatively, WET technologies avoid such nuisances by supplying continuous and stable energy over the air.

Generally speaking, the existing WET technologies could be categorized into three classes based on the key physical mechanisms employed: inductive coupling, magnetic resonant coupling and electromagnetic (EM) radiation. In this paper, we focus on the one based on EM radiation, or radio frequency (RF) enabled WET, which exploits the radiative far-field properties of EM wave, such that receivers could harvest energy remotely from the RF signals radiated by the energy transmitter. RF-enabled WET enjoys many practical advantages over the other two, such as long operating range, low production cost, and small receiver form factor. Besides, energy multicasting could be easily achieved thanks to the broadcasting nature of microwave. Due to the high attenuation of microwave energy over distance, RF-enabled WET is commonly used for supporting low-power devices, such as RFID (radio frequency identification) tags and sensors. However, the recent development of MIMO technology significantly boosts the energy transfer efficiency \cite{2013:Zhang}, which also opens up more potential applications for RF-enabled WET.

\begin{figure}
\centering
  \begin{center}
    \includegraphics[width=0.4\textwidth]{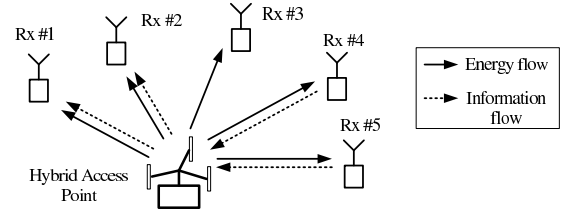}
  \end{center}
  \caption{A generic system model for wireless information and energy transfer.}
  \label{71}
\end{figure}

The technological advances in WET to power wireless devices efficiently open up the potential to build a fully wireless powered communication network (WPCN) without battery replacement. This would significantly reduce the maintenance cost and the frequency of energy outage events due to battery depletion. Another interesting application of WET is to jointly transmit energy and information using the same waveform. Such a design paradigm is commonly referred to as simultaneous wireless information and power transfer (SWIPT), which is proved to be more efficient in spectrum usage than transmitting information and energy in orthogonal time or frequency channels \cite{2013:Zhang,2013:Zhou}.

The concepts of WET, WPCN and SWIPT are illustrated via a generic system model in Fig.~$1$. In the downlink (DL), a hybrid access point (HAP) with stable power supply transmits energy and possibly also information to a set of distributed receivers (Rxs). Meanwhile, the receivers could also transmit information to the HAP in the uplink (UL) using the energy harvested in the DL WET. Then, three canonical operating modes are specified as follows:
\begin{itemize}
  \item WET: energy transfer in the DL only;
  \item SWIPT: energy and information transfer in the DL;
  \item WPCN: energy transfer in the DL and information transfer in the UL.
\end{itemize}
Accordingly, the receivers all perform energy harvesting (EH) in the WET mode, with applications such as charging RFID tags. Additionally, the receivers perform EH in the DL transmission of WPCN mode while sending data in the UL with harvested energy, with applications such as sensor battery charging and data collection in a WSN \cite{2014:Ju1}. For the SWIPT mode, the receivers perform both EH and information decoding (ID) in the DL with the same received signals, each using harvested energy to power its information decoder, e.g., in an energy self-sustainable information broadcast network \cite{2013:Zhang,2013:Zhou}. The system could be extended to a more general network consisting of multiple transmitters and/or receivers with heterogeneous operating modes.

In this paper, we overview the recent advances of WET technologies and their applications to wireless communications. Our objectives are: 1) to introduce RF-enabled WET and existing performance-enhancing techniques; 2) to show the advantages of applying SWIPT to wireless communication networks; 3) to highlight the new design challenges and opportunities in WPCN. Towards this end, we discuss in Sections II to IV the recent advances and future research directions of WET, SWIPT and WPCN, respectively. Finally, we conclude the paper in Section V.

\section{Overview of Wireless Energy Transfer}

\subsection{RF energy receiver and energy beamforming}
\begin{figure}
\centering
  \begin{center}
    \includegraphics[width=0.35\textwidth]{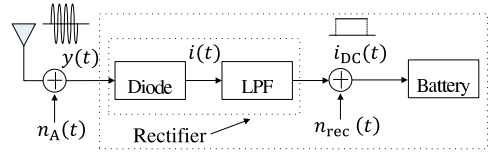}
  \end{center}
  \caption{Structure of an RF energy receiver using a rectifying circuit \cite{2013:Zhou}.}
  \label{72}
\end{figure}

A typical RF energy receiver based on a \emph{rectifying circuit} is given in Fig.~\ref{72}. The receiver converts the received RF signal $y(t)$ to a DC current $i_{\text{DC}}(t)$ through a rectifier, which consists of a Schottky diode and a passive low-pass filter (LPF). The DC current is then used to charge the built-in battery to store the energy. It is shown in \cite{2013:Zhou} that the harvested energy $Q$ per unit symbol time is $Q = \xi  \mathbb{E}\left[|y(t)|^2\right]$, where $\xi \in[0,1]$ denotes the conversion efficiency of the receiver. Notice that the energy contribution from the receiver noise is neglected.

The above energy receiver structure could be easily extended to a receiver with $M>1$ antennas. By the law of energy conservation, it could be assumed that the harvested RF-band power is proportional to that of the received baseband signal. Thus, we consider a discrete baseband MIMO channel
\begin{equation}
\small
\label{9}
\mathbf{y}(n) = \mathbf{H}\mathbf{x}(n) + \mathbf{z}(n),
\end{equation}
where $n$ denotes the symbol index, $\mathbf{x}(n) \in \mathbb{C}^{N\times 1}$ is the baseband transmit signal with zero-mean and covariance $\mathbf{S}\triangleq \mathbb{E}\left[\mathbf{x}(n)\mathbf{x}(n)^H\right]$, $N$ denotes the number of transmit antennas, $\mathbf{H}\in \mathbb{C}^{M\times N}$ represents the complex channel coefficients, and $\mathbf{z}(n)\in \mathbb{C}^{M\times 1}$ is the noise at receiver. Applying the energy receiver in Fig.~$\ref{72}$ to each of the $M$ receive antennas, the total harvested energy is
\begin{equation}
\small
\label{8}
Q = \xi \mathbb{E}\left[||\mathbf{y}(n)||^2\right] \approx \xi \mathbb{E}\left[||\mathbf{Hx(n)}||^2\right] = \xi \text{tr}\left(\mathbf{WS}\right),
\end{equation}
where $\mathbf{W}=\mathbf{H}^H\mathbf{H}$. Given a transmit power constraint $\text{tr}\left(\mathbf{S}\right)\leq P$, the design of the transmit covariance $\mathbf{S}^*$ to maximize the harvested energy is referred to as \emph{energy beamforming} (EB) \cite{2013:Zhang}, where the rank of $\mathbf{S}^*$ indicates the number of beams generated. Then, the optimal EB design is formulated as
\begin{equation}
\small
\label{7}
\begin{aligned}
 \underset{\mathbf{S}}{\text{max}}& & & \xi\text{tr}\left(\mathbf{WS}\right)\\
\text{s. t. }&  & & \text{tr}\left(\mathbf{S}\right)\leq P,\ \ \mathbf{S}\succeq \mathbf{0}.
\end{aligned}
\end{equation}
The optimal solution to (\ref{7}) is shown to be $\mathbf{S}^* = P \cdot \mathbf{v}_1\mathbf{v}_1^H$ \cite{2013:Zhang}, where $\mathbf{v}_1$ corresponds to the eigenvector associated with the largest eigenvalue of $\mathbf{W}$, denoted by $w_1$, and the maximum harvested energy is thus $Q_{\text{max}}=\xi w_1 P$. We see that optimal EB solution $\mathbf{S}^*$ is a rank-one matrix, i.e., generating only one energy beam. Then, the optimal transmit waveform could be expressed as $\mathbf{x}(n)=\sqrt{P}\mathbf{v}_1 s(n)$, where $s(n)$ is an arbitrary random signal with zero mean and unit variance and $\mathbf{v}_1 $ is referred to as the \emph{optimal energy beamforming vector}.

The optimal EB design could be easily extended to energy multicast that maximizes the sum harvested energy at $K\geq 1$ receivers. That is, replacing $\mathbf{W}$ in (\ref{7}) by $\mathbf{\bar{W}} = \sum_{k=1}^K \mathbf{W}_k$, where $\mathbf{W}_k = \mathbf{H}_k^H\mathbf{H}_k$ and $\mathbf{H}_k \in \mathbb{C}^{M_k\times N}$ is the channel from the transmitter to receiver $k$, $k=1,\cdots,K$, each with $M_k$ receive antennas. Similarly, the optimal beamforming solution is also a rank-one matrix. However, this design may lead to severe unfairness in favor of energy receivers closer to the transmitter. Alternatively, we could balance the energy harvesting performance among all receivers by placing additional minimum harvested energy constraint for each receiver, or changing the objective to maximizing the minimum harvested energy of receivers. In these cases, the optimal $\mathbf{S}^*$ is in general no longer a rank-one matrix, indicating that multiple energy beams should be generated \cite{2014:Liu}.

\subsection{Energy beamforming based on imperfect CSIT}
The optimal EB design requires accurate knowledge of channel state information at the transmitter (CSIT). In practice, however, full CSIT is often not available. On one hand, this is because energy receivers, such as the one in Fig.~\ref{72}, in general do not have baseband processing capability to estimate the CSI. On the other hand, accurate channel estimation consumes considerable amount of energy and time, the energy costs of which may offset the performance gain from energy beamforming using more accurate CSI. Nonetheless, we could still improve upon isotropic transmission (setting $\mathbf{S}=\mathbf{I}$) by exploiting the available partial knowledge of CSIT.

When the energy receiver is unable to estimate CSI, \cite{2014:Xu1} proposes an EB design method where the transmitter adjusts its energy beamforming vector based on only one-bit feedback sent by the energy receiver(s), indicating the increase or decrease of the harvested energy in the current iteration compared to the last. The proposed method converges to the optimal EB solution as the iterations proceed. For a receiver with more complex signal processing capabilities to estimate the CSI, \cite{2014:Yang} proposes adaptive online and offline methods to maximize the harvested energy, where the receiver uses the current noisy channel estimate to decide whether to continue to perform channel estimation or feed back to the transmitter to initiate energy transfer. Besides, \cite{2014:Zeng} exploits the UL/DL channel reciprocity by allowing the transmitter to estimate the DL channel from the pilot signals sent by the receiver. By considering the energy and time consumed on pilot signal transmission, it maximizes the harvested energy by selecting an optimal subset of receiver antennas to transmit pilot signals.

\emph{Challenges and opportunities:} Due to receiver hardware constraints, accurate CSIT is often not available in the context of WET. The EB design based on one-bit feedback proposed in \cite{2014:Yang} is attractive. However, it may take many iterations to converge to an optimal solution, especially when the dimension of the MIMO channel is large. It is therefore worth studying channel learning and EB design schemes with faster convergence speed under limited feedback to improve WET performance, taking into account the tradeoff involved due to the increase in the amount of feedback information.

\begin{figure}
\centering
  \begin{center}
    \includegraphics[width=0.4\textwidth]{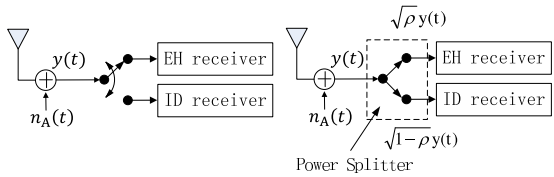}
  \end{center}
  \caption{Two practical designs for the co-located energy and information receivers: time switching (left) and power splitting (right) \cite{2013:Zhang}.}
  \label{73}
\end{figure}

\section{Simultaneous Wireless Information and Power Transfer}
\subsection{Practical SWIPT receiver structure}
The idea to simultaneously transmit information and energy using the same RF waveform was first proposed by Varshney \cite{2008:Varshney}, which characterizes the \emph{rate-energy (R-E) tradeoff} in a discrete memoryless channel, defined as the maximum achievable data rate under a received energy constraint. The study of R-E tradeoff is later extended to frequency selective channel \cite{2010:Grover}, multiple access and multi-hop channels \cite{2012:Fouladgar}, and two-way channels \cite{2013:Popovski}. However, all the works above are based on the same ideal assumption that the receiver is able to decode information and harvest energy from the same signal, which is not realizable yet due to the practical circuit limitations.

The first practical receiver structure that enables SWIPT is proposed in \cite{2013:Zhang}. As shown in Fig.~$\ref{73}$, it consists of co-located ID and EH receivers, where the ID receiver is a conventional information decoder and the EH receiver's structure follows that in Fig.~$\ref{72}$. Two practical signal separation schemes are considered: \emph{time switching} (TS) and \emph{power splitting} (PS). For the TS scheme, the transmitter divides the transmission block into two orthogonal time slots, one for transferring power and the other for transmitting data. Accordingly, the receiver switches its operations periodically between harvesting energy and decoding information between the two time slots. Different R-E tradeoffs could be achieved by varying the length of energy transfer slot. On the other hand, the PS scheme splits the received signal into two streams, where one stream with power ratio $0\leq \rho\leq 1$ is used for EH and the other with power ratio $\left(1- \rho\right)$ is used for ID. Different R-E tradeoffs are therefore achievable by adjusting the value of $\rho$.

\begin{figure}
\centering
  \begin{center}
    \includegraphics[width=0.4\textwidth]{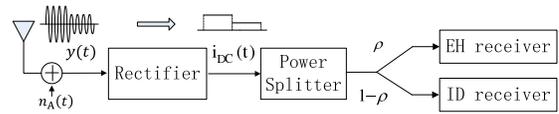}
  \end{center}
  \caption{Structure of integrated information and energy receiver \cite{2013:Zhou}.}
  \label{74}
\end{figure}

Another practical integrated receiver (IntRx) structure is proposed by \cite{2013:Zhou} and illustrated in Fig.~$\ref{74}$. Unlike the separate receiver (SepRx) in Fig.~$\ref{73}$ that splits the signal at the RF band, the IntRx combines the RF front-ends of ID and EH receivers and splits the signal after converting it into DC current. Similar TS and PS methods could then be applied to separate the DC current to perform EH and ID, respectively. IntRx uses a passive rectifier for RF-to-baseband conversion, which saves the circuit power consumed by the active mixer used in the information decoder of SepRx. However, the ID receiver needs to perform noncoherent detection from the baseband signal. In this case, conventional phase-amplitude modulation (PAM) must be replaced by a capacity-reduced \emph{energy modulation}, where information is only encoded in the power of the input signal. In a SISO AWGN channel, \cite{2013:Zhou} studies the R-E tradeoffs of both SepRx and IntRx using TS and PS methods. In general, IntRx is superior than SepRx when more harvested energy is required, while SepRx outperforms IntRx in the high-data-rate region. In most of the existing studies, the SepRx is considered due to its simplicity in hardware and encoding/decoding design. Unless otherwise stated, we only consider the SepRx structure in the following discussions.

Channel fading and interference are two major challenges to the transceiver design of wireless communication. However, they result in quite different performance degradation for ID and EH in SWIPT. While deep channel fading degrades the performance of both ID and EH, strong interference is only harmful to ID but in fact helpful to increase the harvested energy for EH receiver. Within this context, the optimal receiver TS strategies in a SISO fading channel under co-channel interference are studied in \cite{2013:Liu1} to maximize the average data rate under an average EH constraint. In particular, it shows that the receiver should perform ID when the received signal (information and interference) is relatively weak and the SNR is sufficiently high, and EH otherwise. Intuitively, this is because the ID receiver gains less than the EH receiver when the interference is strong (harmful for ID but helpful for EH), and when the channel is in good condition (logarithmic increase in rate for ID but linear increase for EH). When CSIT is available, the transmitter could also adapt its transmit power to the channel state, e.g., it does not transmit under deep channel fading. Besides, the optimal PS strategies in SISO/SIMO fading channels are derived in \cite{2013:Liu2}.

The application of MIMO technology could significantly enhance the transfer efficiency of both energy (energy beamforming) and information (spatial multiplexing). SWIPT in MIMO AWGN channels are first studied in \cite{2013:Zhang}, which derives the achievable R-E regions of various schemes. In particular, the use of multiple antennas at the receiver enables a low-complexity
implementation for PS: \emph{antenna switching} (AS), where a subset of antennas is used for EH ($\rho= 1$), while the rest are for ID ($\rho= 0$). AS reduces the hardware complexity of PS as it does not need a power splitter for each antenna. Instead, it simply connects an antenna to either an ID receiver or an EH receiver. However, the saving also comes at a cost of a smaller R-E region. The application of AS is also investigated in SIMO fading channels in \cite{2013:Liu2}, which proposes both optimal and reduced-complexity suboptimal algorithms to determine the subset of antennas used for ID and EH.

\subsection{Network applications of SWIPT}
With the SWIPT receiver structures introduced above, the applications of SWIPT have been studied in various network settings. In broadcasting channels, \cite{2013:Zhang} first characterizes the R-E tradeoff for a transmitter transferring energy and information to two separated/co-located ID and EH receivers. \cite{2014:Xu} and \cite{2014:Shi} extend the optimal beamforming designs to general broadcasting channels, consisting of multiple separated and co-located ID and EH receivers, respectively. Besides, resource allocations of SWIPT in OFDM-based multiuser systems are studied in \cite{2013:Ng1,2014:Zhou,2013:Huang}. When perfect CSIT is not available, \cite{2012:Tao} proposes a robust beamforming design and \cite{2014:Ju} proposes a random beamforming method to improve the R-E tradeoff. SWIPT in relay channels is studied in \cite{2012:Krikidis,2013:Nasir}, where \cite{2013:Nasir} derives the achievable throughput when the relay adopts TS or PS receiving strategies. SWIPT in interference channel is investigated in \cite{2012:Shen,2013:Park,2014:Timotheou,2014:Lee}. In particular, \cite{2014:Lee} shows that the R-E tradeoff could be significantly improved by jointly designing the energy signals across all the transmitters. In addition, \cite{2014:Liu} and \cite{2014:Xing} consider an interesting application scenario where information secrecy is incorporated in the design of SWIPT.

\emph{Challenges and opportunities:} Most of the current works consider the SepRx structure. In practice, however, IntRx outperforms SepRx in the high harvested-energy region, and may be strictly better than SepRx when the circuit power is high \cite{2013:Zhou}. The major difficulty of applying IntRx lies in determining the capacity of the end-to-end channel and the corresponding optimal input distribution. Alternatively, practical energy modulation schemes could be investigated in the future, with some initial attempts in \cite{2013:Zhou}. Besides, the fundamental limits of SWIPT, not constrained by the receiver structures introduced in this paper, are still open.

\section{Wireless Powered Communication Network}
\subsection{DL/UL designs in a single-cell WPCN}
WET enables the replenishment of the energy to power wireless devices without disruption to the use. This also leads to a fundamental shift in design principles compared to conventional systems with stringent battery lifetime constraints: instead of minimizing energy consumption as the most critical objective for battery-powered networks, in a WPCN the objective becomes balancing the energy supply (using WET) and consumption (on data transmission), to optimize the network performance. In the most simple setting of WPCN in Fig.~\ref{71}, the wireless devices transmit data in the UL using the energy harvested from the DL transmission. The difficulties of system design in WPCN mainly lie in twofold: 1) the DL/UL energy and information transmissions are asymmetric in nature; 2) the DL/UL transmissions are coupled by the energy constraints of the wireless devices. Therefore, the DL/UL transmissions must be jointly designed, which inevitably introduces a tradeoff in resource allocation.

\begin{figure}
\centering
  \begin{center}
    \includegraphics[width=0.4\textwidth]{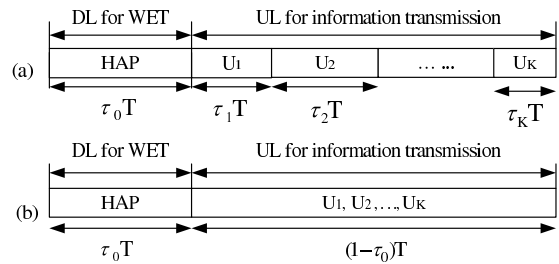}
  \end{center}
  \caption{Two harvest-then-transmit protocols for WPCN: (a) TDMA-based (above) \cite{2014:Ju1} and (b) SDMA-based (below) \cite{2014:Liu2}. $T$ is the duration of a transmission block. $\tau_0$ is portion of time used for DL energy transfer. $\tau_k$ is the portion of time used for UL data transmission by user $U_k$ in TDMA-based protocol with $\tau_0 + \tau_1+\cdots +\tau_K =1$.}
  \label{75}
\end{figure}

To address the problem, \cite{2014:Ju1} proposes a \emph{harvest-then-transmit} protocol in a WPCN with a single-antenna HAP. As illustrated in Fig.~\ref{75}(a), multiple users first harvest energy from the DL WET and then transmit data in the UL in a TDMA manner. By optimizing the durations of $\tau_k$'s, it first considers maximizing the UL sum throughput. However, such design will induce a serious \emph{doubly near-far problem}, where users far away from the transmitter achieve very low throughput, as they suffer from both low harvested energy in DL and high transmission power consumption in UL. The problem is resolved by maximizing the worst user throughput in the network. This work is later extended in \cite{2014:Ju2} and \cite{2014:Kang} to use a full-duplex HAP to transmit energy and receive user data simultaneously, where the key challenge is to overcome the self-interference caused by the full-duplex operation.

Instead of using TDMA for multiple access, \cite{2014:Liu2} extends the single-antenna HAP in \cite{2014:Ju1} to a multi-antenna HAP that enables spectrum efficient SDMA (space-division-multiple-access). The operation of the system is illustrated in Fig.~\ref{75}(b). To tackle the doubly near-far problem, it maximizes the minimum UL throughput among all users by jointly designing the UL/DL slot lengths, DL energy beamforming and multi-user detection (MUD) in the UL. With sufficiently large number of antennas at the HAP, \cite{2014:Yang1} studies the design of WPCN using massive MIMO technology. The channel properties of massive MIMO significantly simplify the designs of EB and MUD as in \cite{2014:Liu2}. In particular, it includes the consideration of channel estimation in the system design, where the lengths of channel estimation (UL), energy transfer (DL), and information transmission (UL) are jointly optimized. In addition, \cite{2014:Ju3} proposes a user cooperation method to overcome the doubly near-far problem, where a user close to the HAP helps relay the message of a far user to increase its end-to-end throughput.

\subsection{Multi-cell network capacity}
The design focus in the above single-cell WPCNs is joint resource allocation in the DL/UL transmissions. When we extend the study to multi-cell WPCN, global resource allocation is very complicated and often intractable. To obtain more insight into the design of multi-cell WPCN, the method of \emph{stochastic geometry} is widely used to study the scaling laws of network capacity as a function of system parameters.

In a multi-cell cellular network, \cite{2014:Huang1} proposes to install dedicated microwave energy transferring nodes, named power beacons (PBs), to support the mobile users' data transmission to the base stations (BSs). Transmission outage occurs due to energy depletion, signal attenuation or channel fading. Intuitively, to maintain a relatively low outage probability, the BSs and PBs should be densely deployed and transmit with high power. Using a stochastic geometry model, the functional relationships between BS and PB densities as well as their transmission power, are derived to achieve a prescribed transmission outage probability. Besides, it also shows the performance gains brought by large battery capacity and directional energy beamforming. The performance of a wireless powered multi-cell cognitive radio network is studied in \cite{2014:Lee1}. It assumes that a secondary transmitter could harvest energy from a primary transmitter if they are close enough. Besides, the secondary transmitter could transmit to its receiver as long as it is not in the guard zone of any primary transmitter. The outage probabilities of both primary and secondary transmissions are derived based on a stochastic geometry model, and the secondary network throughput is maximized under the prescribed outage constraints.

\emph{Challenges and opportunities:} WPCN opens up a new design paradigm for wireless networks, where many designs of PHY, MAC, and network layers in conventional wireless networks need to be revisited. A promising research direction is to combine SWIPT in the DL and data transmission in the UL via an information/energy full-duplex approach. Another important topic is efficient channel estimation scheme, as now it affects both the UL data transmission and DL energy transfer efficiency. Besides, the prototyping and evaluation of WPCN in practice are also of significant importance.

\section{Conclusions}
In this paper, we have provided an overview of the recent advances in joint energy and information transfer. We showed that RF-enabled WET technology could be easily applied in wireless communication networks with simple and inexpensive transceiver structures. The performance gain of joint energy and information transfer was demonstrated by studying two design paradigms: SWIPT and WPCN. We envision that the future wireless system is a mixture of energy and information transfer, where the architecture and system design must capture the interplay between the energy and information characteristics in the network.

% that's all folks
\end{document}